\documentclass[onecolumn,preprintnumbers,aps,12pt,longbibliography,showkeys]{revtex4-2}

\usepackage[utf8]{inputenc} \usepackage[T1]{fontenc}
\usepackage{times}
\usepackage{amsmath}
\usepackage{bm}
\usepackage{amssymb,latexsym,mathrsfs}
\usepackage{url}
\usepackage{color}
\usepackage{tikz}    
\usepackage{xcolor}  

\usepackage[sf,pagestyles]{titlesec}
\titleformat{\section} {\normalfont\sffamily\bfseries} {\thesection}{1em}{}

\usepackage{psfrag} 
\usepackage{xurl}
\usepackage[bookmarks=false,breaklinks,bookmarksnumbered=true]{hyperref}
\usepackage[hyphenbreaks]{breakurl}

\usepackage[singlelinecheck=false,skip=-2mm,labelfont={bf,sf},font=small]{caption}

\definecolor{royalbluecs}{rgb}{0.2, 0.2, 1}

\hypersetup{linktoc=all, 
    colorlinks=true, linkcolor={royalbluecs},
    citecolor={royalbluecs}, urlcolor={royalbluecs}}

\definecolor{lime}{HTML}{A6CE39}
\DeclareRobustCommand{\orcidicon}{%
	\begin{tikzpicture}
	\draw[lime, fill=lime] (0,0) 
	circle [radius=0.16] 
	node[white] {{\fontfamily{qag}\selectfont \tiny ID}};
	\draw[white, fill=white] (-0.0625,0.095) 
	circle [radius=0.007];
	\end{tikzpicture}
	\hspace{-2mm}
}

\foreach \x in {A, ..., Z}{%
	\expandafter\xdef\csname orcid\x\endcsname{\noexpand\href{https://orcid.org/\csname orcidauthor\x\endcsname}{\noexpand\orcidicon}}
}

\makeatletter
\def\incepsfig{\@ifnextchar[{\@incepsf}{\@incepsf[tp]}}
\def\@incepsf[#1]#2#3#4{\@ifnextchar[{\@incepsfp[#1]{#2}{#3}{#4}}{\@incepsfp[#1]{#2}{#3}{#4}[]}}
\def\@incepsfp[#1]#2#3#4[#5]{%
\begin{figure*}[#1]%
  \captionsetup{width=140mm}
  \begin{center}%
	  \scalebox{#3}{\hss\begin{psfrags}#5{\includegraphics{#2.eps}}\end{psfrags}\hss}%
  \end{center}%
  \caption{#4}\label{#2}
  \vspace{-4mm}
\end{figure*}}
\makeatother

\def\figureref#1{{{Figure~\ref{#1}}}} 

\long\def\comment#1{} 

\usepackage{etoolbox}
\makeatletter
\patchcmd{\frontmatter@RRAP@format}{(}{}{}{}
\patchcmd{\frontmatter@RRAP@format}{)}{}{}{}
\renewcommand\Dated@name{} 
\makeatother

\normalsize

\begin{document}

\title{\textsf{From the Bronshtein cube of limits \\ to \\ the degrees of freedom of relativistic quantum gravity \\ \ }}

\author{\ \\ 
Christoph Schiller\footnote{{Motion Mountain Research, 81827 Munich, Germany, 
\email{cs@motionmountain.net}},
ORCID 0000-0002-8188-6282.} \orcidA{} \\
\ \\ }


\begin{abstract} 
\noindent \ \hfil \par \noindent {\bf\textsf{Abstract}}\par\noindent
It is argued that the quadruple gravitational constant $4G$ can be seen as a fundamental limit of nature. The limit holds across all gravitational systems and distinguishes bound from unbound systems. Including the maximum force $c^4/4G$ allows extending the Bronshtein cube of physical theories to a cube of limits. Every theory of physics refining Galilean physics -- universal gravitation, special relativity, general relativity, quantum theory and quantum field theory -- is defined by one fundamental limit. As a result, also relativistic quantum gravity is defined by a limit: the minimum length in nature. The minimum length is used to deduce the Planck-scale structure of space. Numerous options are eliminated. Then, the minimum length is used to deduce the main properties of the common constituents that make up space and particles. 
\end{abstract}

\bigskip

\bigskip

\hbox to \textwidth{\hss\emph{\ }\hss}

\hbox to \textwidth{\hss\emph{Essay written for the Gravity Research Foundation 2023 Awards for Essays on Gravitation.}\hss} 

\bigskip

\bigskip

\bigskip

\bigskip

\date{February 2023 \\ } 

\bigskip

\bigskip

\maketitle
\newpage

\section{The Bronshtein cube}

\noindent 
It is common to attribute to Matvei Petrovich Bronshtein, born in 1906 in Vinnytsia in Ukraine, the physics cube that is based on the speed of light $c$, the quantum of action $\hbar$ and the gravitational constant $G$ \cite{bronsteinbio,bronstein1933,bronstein1934}. 
The physics cube has regularly been used to illustrate the relation between the main physical theories \cite{Okun:2001rd,Okun:2002gnu,padmanabhan2015grand,Oriti:2018tym}.

In the following, it is first argued that Bronshtein's usual physics cube can be extended to a cube of \textit{limits}, illustrated in \figureref{Bronshtein-V3}.
Since their inception, special relativity and quantum theory are based on the invariant maximum speed $c$ and the invariant minimum action $\hbar$. 
In the past decades, it became clear that general relativity can be based on the invariant maximum force ${c^4}/{4G}$, which is realized on gravitational horizons \cite{gibbons,s1,csmax,MYNEWPRD,GRF}. 
Below, the approach is completed by showing that also classical gravity can be based on an invariant limit, namely the quadruple gravitational constant $4G$, which is realized in parabolic gravitational motion. 
In this way, each part of fundamental physics is defined using an invariant limit principle. 
In particular, it follows that relativistic quantum gravity is based on the minimum length -- or on an equivalent limit.

It is then argued that the minimum length at the top of Bronshtein's limit cube implies a unique and common description for space and particles that is based on fluctuating filiform constituents. 
The Bronshtein cube of limits thus appears to determine the theory of relativistic quantum gravity.

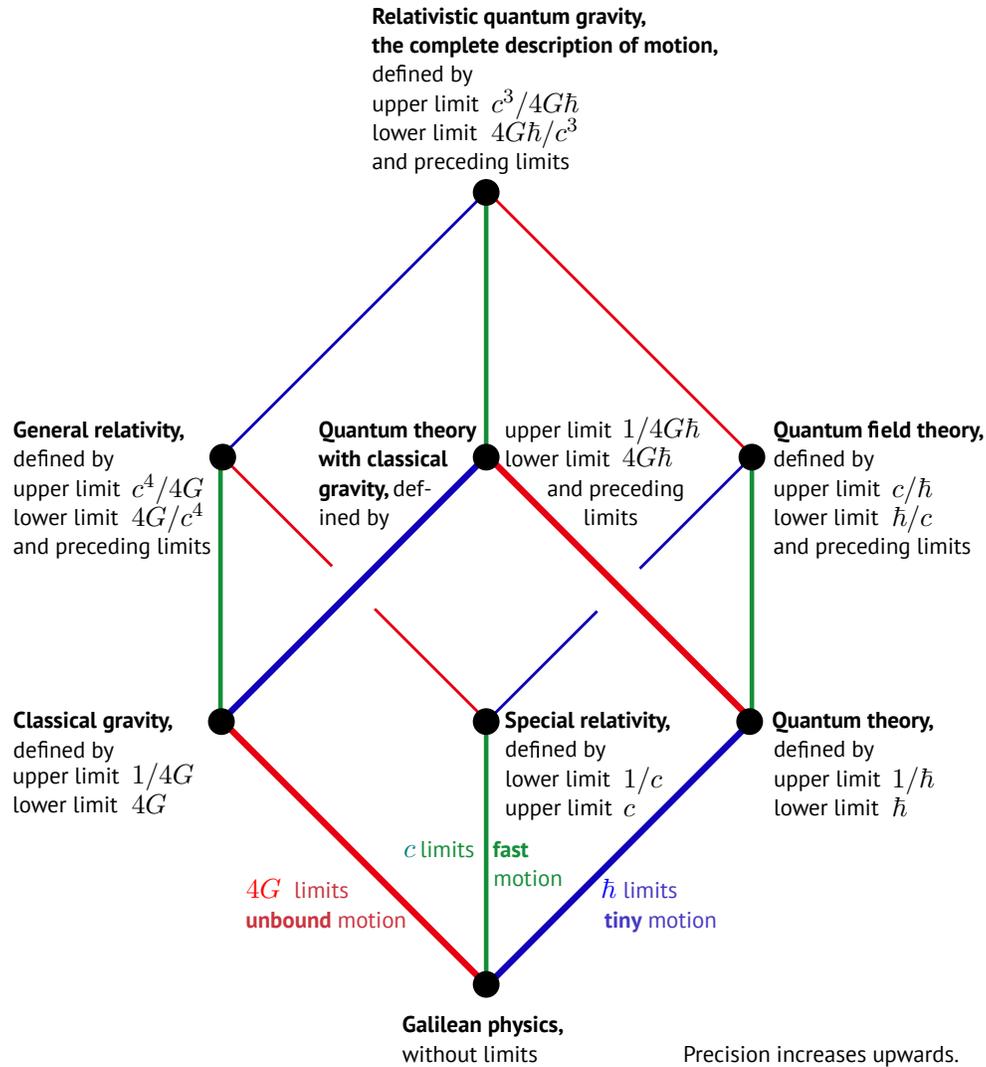
\incepsfig{Bronshtein-V3}{0.95}{The Bronshtein \emph{limit} cube of physical theories is illustrated, in which each modern theory of physics is described by a limit.
Following the lines towards the top increases the precision of the descriptions  -- and their fascination.
}[%
\psfrag{Z}{\textcolor{teal}{\small $c\,$}}%
\psfrag{Y}{\textcolor{red}{\small $4G$}}%
\psfrag{HB}{\textcolor{blue}{\small $\,\,\hbar$}}%
\psfrag{UL1}{\small $1/4G$}%
\psfrag{LL1}{\small $4G$}%
\psfrag{UL2}{\small $c$}%
\psfrag{LL2}{\small $1/c$}%
\psfrag{UL3}{\small $1/\hbar$}%
\psfrag{LL3}{\small $\hbar$}%
\psfrag{UL4}{\small $c^4/4G$}%
\psfrag{LL4}{\small $4G/c^4$}%
\psfrag{UL5}{\small $1/4G\hbar$}%
\psfrag{LL5}{\small $4G\hbar$}%
\psfrag{UL6}{\small $c/\hbar$}%
\psfrag{LL6}{\small $\hbar/c$}%
\psfrag{UL7}{\small $c^3/4G\hbar$}%
\psfrag{LL7}{\small $4G\hbar/c^3$}%
]
\section{Maximum force} 

\noindent In 1973, Elizabeth Rauscher, followed by many others, discovered that general relativity \emph{implies} a local maximum force $c^4/G$
\cite{rauscher,treder,10.2307/24530850,sab,massa,Kostro:1999ue,gibbons,s1,csmax,csmax2,barrow1,abo,ong2,Bolotin:2016roy,maxlum,Ong:2018xna,barrow2,barrow3,b1,atazadeh,jowsey,faraoni,newcs,Faraoni:2021sep,Cao:2021mwx,MYNEWPRD,Dadhich:2022yuk,hogan,barrow1,ong2,abo,Ong:2018xna,barrow2,barrow3,b1,MYNEWPRD,Ong:2018xna,Gurzadyan:2021hgh,michael2,GRF,loeb2022three}.
In 2002, Gibbons, and the present author in 2003, included the factor $1/4$ and showed that in 3+1 dimensions the value of the force -- the change of momentum of a system with time -- produced by black holes on any system at a point is never larger than the maximum value $c^4/4G$ \cite{gibbons,s1}.

The (local) maximum force value $F_{\rm max}={c^4}/{4G} \approx 3.0 \cdot 10^{43} \rm \,N$ is due to the maximum energy per distance ratio appearing in general relativity.
For example, for a Schwarzschild black hole, the ratio between the energy $Mc^2$ and its diameter $D=4GM/c^2$ is given by the maximum force value, independently of the size and mass of the black hole.  
Also the force on a test mass that is lowered with a rope towards a gravitational horizon -- whether charged, rotating or both -- never exceeds the force limit, \emph{as long as} the minimum size of the test mass is taken into account.  
All apparent counter-arguments and counter-examples to maximum force disappear when explored in detail \cite{jowsey,faraoni,newcs,Faraoni:2021sep,MYNEWPRD}.

As a note, maximum force $c^4/4G$ \emph{implies} Einstein's field equations of general relativity \cite{gibbons,s1,csmax,MYNEWPRD,GRF}.
This result, only valid in 3+1 dimensions, can be reached in two ways:
it can be deduced from the elastic properties of space-time implied by maximum force, and it can be deduced from $c^4/4G$ with the help of the first law of black hole horizons.
As a result of these deductions, the (local) maximum force limit can be seen as the defining \emph{principle} of general relativity.  
The situation resembles special relativity, for which the (local) maximum speed limit $c$ can be seen as the defining principle.

The maximum force limit is not the only option to describe general relativity.
Other limit quantities that combine $c$ and $G$, such as maximum power $c^5/4G$ \cite{sciama1,MYNEWPRD,lastsivaram,massa,hogan,Kostro:2000gw,Cardoso:2018nkg,Ong:2018xna,abo,Gurzadyan:2021hgh},
maximum mass flow rate $c^3/4G$ \cite{MYNEWPRD,Cao:2021mwx}, or the maximum mass per length ratio $c^2/4G$ \cite{thorne,Hod:2020yub,cshoop} can also be taken as principles of relativistic gravitation. 
All these limits values are achieved only by black holes. 
A similar situation arises in special relativity, where, e.g., either $c$ or $c^2$ can be taken as defining limit. 
All these limits agree with all experiments.

In short, special relativity can be deduced from the principle of maximum speed $c$;
general relativity can be described by adding 
maximum force $c^4/4G$.
A question arises: can classical inverse square gravity also be deduced from a limit?

\section{The quadruple gravitational constant as a limit distinguishing unbound from bound particles}
\label{bdef}

\noindent The gravitational constant $G$ is defined as the 
constant appearing in the gravitational acceleration of a small mass, or particle, at a distance $r$ to a large mass $M$:
\begin{equation}
	 a= \frac{GM}{r^2} \;\;.
	 \label{eq1}
\end{equation}
This expression, independent of the value of the small mass, is due to the work of Hooke, Newton and Cavendish. 
Because the expression unified sublunar and translunar observations, the inverse square dependence of classical gravity is often called \emph{universal} gravity.
%
Rewriting the equation using the diameter $d=2r$ yields
\begin{equation}
	 a= \frac{4GM}{d^2} \;\;.
	 \label{eq2}
\end{equation}
The expression implies that a small mass or particle is {unbound} from a large mass $M$ if its kinetic energy is larger than the gravitational potential energy. 
In other terms, a particle is \emph{unbound} if the double centre-to-centre distance $d=2r$ and the speed $v$ obey
\begin{equation}
	 \frac{d v^2}{M} \geq 4G \approx 2.7 \cdot 10^{-10}\,\rm m^3/kg\,s^2   \;\;.
	 \label{eq3}
\end{equation}
The constant $4G$ thus describes the difference between \emph{unbound} and \emph{bound} particles near a mass $M$.
If a particle has a product ${d v^2}/{M}$ that is larger than $4G$, it is unbound; otherwise, it is bound.
For example, a rocket `orbiting' Earth with diameter $d$, flying faster than the escape velocity $v= \sqrt{4GM/d} $, is unbound.
In contrast, a stone on the ground is bound to Earth.

Equivalently, a particle is \emph{unbound} from a large mass $M$ if
its acceleration obeys ${ad^2}/{M} \geq {4G}$.
Alternatively, a particle near a mass $M$ is \textit{unbound} if it obeys
\begin{equation}
	{\rho T^2} \leq \frac{1}{4G}  \approx 3.7\cdot 10^{9}\,\rm kg\, s^2/m^3 
 \;\;.
	\label{eq5}
\end{equation}
In this expression, the `effective' particle cycle time $T=d/v$ around the central mass and the `effective' density $\rho=M/d^3$ are used.
In all other cases, the particle is gravitationally bound.



When a particle approaches the limit $4G$ from below, its orbital time increases without limit, 
until parabolic motion is reached. 
Approaching the limit $4G$ from
above leads from \emph{unbound} hyperbolic to parabolic motion, for which gravitational potential energy and kinetic energy are exactly equal.
Crossing the limit \emph{bounds} the particle or test mass to the attracting mass $M$.

In short, for a gravitationally \emph{unbound} particle, $4G$ is the \emph{limit} value of any product of quantities containing the nearby mass $M$. 
All experiments in the solar system confirm the result.

As shown in equation (\ref{eq5}), the value $1/4G$ is the \emph{largest} possible value for effective density times time squared that can arise in a gravitationally \emph{unbound} system.
This limit can also be applied to the universe as a whole.
The quantity $1/4G$ should limit the product $\varrho\, T_H^2$ of the matter 
density in the universe and the (Hubble) time squared. 
Present data on the energy density and the Hubble time \cite{Planck:2018vyg} indeed confirm that, within measurement errors, the limit $1/4G$ is realized by the universe.
The mass-energy in the universe, seen at a large scale, is generally at the limit between being bound and unbound -- as expected. 
In short, the limit $4G$ also holds in cosmology.

\section{Testing the limit $\bm{4G}$ in rotating galaxies}

\noindent
At present, the validity of the limit $4G$ is in discussion for one class of physical systems.
The rotation of outer stars orbiting galaxy centres is an intense topic of research.
In almost all galaxies the most distant stars are measured to rotate \emph{faster} than predicted from inverse square gravity with the estimated central mass values.
Similar results are found for globular clusters and for galaxy groups.
Different explanations have been proposed for these observations.

In the most common explanation, the deviation is explained with yet unobserved \textit{(cold) dark matter} \cite{Planck:2018vyg}.
The explanation postulates that stars (or galaxies) at the outer edge remain bound because the actual mass of galaxies is larger than the luminous mass, because of additional dark matter. This approach retains the limit $4G$.

In contrast, some researchers explore \emph{modifications} of the inverse square dependence of gravitation \cite{McGaugh:2020ppt}.
They postulate that at distances that would lead to accelerations smaller than an observed constant $a_0 \approx 1.2 \cdot 10^{-10}\,\rm m/s^2$, the actual acceleration due to gravitation is \emph{larger} than the one predicted by inverse square gravity, and thus outer stars (or galaxies) remain bound even if their speeds are larger than the conventional escape velocity.
This approach predicted the observed baryonic Tully-Fisher relation between galaxy mass and asymptotic rotation velocity. 
Some versions of modified gravity put into question the limit $4G$.

It might be that the observational constant $a_0$ and the baryonic Tully-Fisher relation are due to some quantum effect on cosmological scale \cite{verlinde2017emergent}, that the Weyl metric explains the rotation curves \cite{immirzi1,immirzi2}, or that measurements are not interpreted correctly.
In these cases, the limit $4G$ would remain valid. 
Future research will show which explanation is correct.

In short, so far, there is no definite observation contradicting the limit $4G$ for unbound motion.

\section{Properties of the limit $4G$}

\noindent
The factor 4 in the limit for gravitationally bound motion arises because of the historical preference to use the radius instead of the diameter.
High-precision experiments confirm equations (\ref{eq3}) and (\ref{eq5}), as well as the expression ofr teh escape velocity.
These expressions allow stating: the limit $4G$ contains and implies the inverse square law of universal gravitation.
The limit $4G$ thus \emph{implies} classical gravity, in the same way that the limit $c$ implies special relativity or the limit $c^4/4G$ implies general relativity. 



The limit $4G$ -- like the fundamental limits $\hbar$, $c$, $c^2/4G$ and $c^4/4G$ --  is \emph{invariant}: it is independent of the observer. 
Neither relativistic boosts nor other coordinate transformations of any type change the limit value. 
Like the other limits, also the limit $4G$ applies only to \emph{real and free} particles. 
The limit $4G$, like all other invariant limits of nature, is also needed to \textit{define units} of measurements. 

In short, the expression $4G$ is a fundamental, invariant limit principle of nature.
This result allows a striking formulation of Bronshtein's physics cube.


\section{The Bronshtein limit cube of physical theories}

\noindent 
As illustrated in \figureref{Bronshtein-V3}, the above results for universal gravity and for general relativity allow defining a \emph{limit} value at every corner of the physics cube -- except for Galilean physics. 
The origin of this possibility is the following summary of modern physics:

\smallskip

\textbullet\  Universal gravity is equivalent to $ dv^2/M \geq 4G$ for unbound masses. 

\textbullet\  Special relativity is equivalent to $v \leq c$ for physical systems.

\textbullet\  General relativity is equivalent to $F \leq c^4/4G$ or $m/d \leq c^2/4G$ for physical systems.

\textbullet\  Quantum theory is equivalent to $W \geq \hbar$ for all physical systems.

\textbullet\  Quantum field theory is equivalent to $ ml \geq \hbar/c$ for real, i.e., physical particles.  

\textbullet\  Non-relativistic quantum gravity is equivalent to free particles with $ Av^3 \geq 4G\hbar$.  

\textbullet\  Relativistic quantum gravity is equivalent to $ l^2 \geq {4G\hbar/c^3}$ for physical systems.  

\smallskip

\noindent 
In these expressions, $l$ is length, $W$ is action, and $A$ is area.
Each of the seven limit expressions is associated with one of the upper corners of Bronshtein's physics cube, as illustrated in \figureref{Bronshtein-V3}. 
The choice of the limit for every field of physics is not unique; at each corner of the cube,
the nonzero exponents of $4G$, $c$ and $\hbar$ can be changed within certain boundaries.
These choices allow assigning several equivalent upper or lower limits to each corner of the cube.
If desired, the limit for entropy, defined by the Boltzmann constant $k$ \emph{can} be added to the description of nature, yielding a hypercube \cite{Okun:2001rd,Oriti:2018tym,ninelines}.

Several properties of the Bronshtein \emph{limit} cube are worth pointing out.
First, apart from the correcting factor 4, all limits are Planck limits.
In other terms, one can say:  \textit{the (corrected) Planck limits define modern physics.} 
Each Planck limit is a \textit{principle} for the corresponding theory.

Each Planck limit is deduced from experiments.  
\textit{No} experiment ever performed contradicted any of these limits. 
In other terms, there is no evidence for trans-Planckian physics from any observation.
The  Bronshtein limit cube implies that there is \emph{no} physics beyond general relativity, even for the strongest gravitational fields. 
The cube implies that there is \emph{no} physics beyond quantum field theory, even for the smallest scales or the highest energies.
The cube further implies that there is \emph{no} physics beyond quantum theory in classical gravity. 
This agrees with all experiments ever performed.

Whenever a limit $c$, $4G$ or $\hbar$ is \textit{added} to a 
given description of motion, a \emph{more precise} description is obtained.
Because relativistic quantum gravity takes into account \emph{all} limits, it is the most precise description of fundamental physics. Thus, relativistic quantum gravity is the complete and final theory of physics.

In particular, the combination of all limits implies a statement on length.  Using the expression for action $W = F l t = F l^2/v$, and inserting the limits for force $F$, speed $v$ and action $W$, one finds that length values are limited by $l \geq \sqrt{4G\hbar/c^3}$, or twice the Planck length.
This result has been derived in many ways in the past \cite{mead,garay, maggiore1993generalized, sabinelength, tawfik2015review}. 
The smallest length limits both the observable values and the achievable measurement precision.
Many other, equivalent limits -- such as the minimum area, the minimum time, the minimum volume, the maximum acceleration or the maximum mass density -- can also be deduced for the top corner of the physics cube.

In short, each field of fundamental physics can be defined with a Planck limit principle.
This characterization is also useful for teaching:
learning physics starts at the bottom of the Bronshtein cube, where \textit{no} limits are assumed, and proceeds towards the top, where \textit{all} physical observables are limited by the corrected Planck values. 
The agreement with experiments, the simplicity, and the explanatory power of the fundamental limits at each corner of the physics cube are fascinating.
The fascination is especially intense at the top corner.

\section{Relativistic quantum gravity and its predictions}

\noindent 
In the Bronshtein cube of limits, any path towards the summit leads to the theory of relativistic quantum gravity.
This implies that the minimum length limit $\sqrt{4 G \hbar/ c^3}$ -- or any equivalent quantum gravity limit --  
\emph{alone by itself}, completely describes and {implies} relativistic quantum gravity.

Every limit that can be used to define the top of the physics cube -- from minimum length to minimum time, minimum area, minimum volume, maximum acceleration or  maximum mass density -- is \textit{inaccessible} to experiments.
Therefore, the Bronshtein limit cube suggests that relativistic quantum gravity is already known in all its observable effects:
the only observable effects of relativistic quantum gravity are predicted to be general relativity, quantum field theory with the standard model, and quantum theory in classical gravity. 
These are the theories for the three corners below the top of the cube.

In addition, minimum length, together with its various equivalent limits of nature at the uppermost corner, leads to the following conclusions about relativistic quantum gravity:

\textbullet\       At the Planck scale, minimum length and minimum time imply that space-time is \textit{not a manifold} and thus has no defined dimensionality. In contrast, at macroscopic scale -- i.e., at all scales larger than the Planck scale -- space and time are {effectively} continuous. 
Space and time emerge from the Planck limits as macroscopic approximations. 
At scales larger than the Planck scale, space-time is \textit{effectively}, i.e., \textit{approximately} a 3+1 dimensional manifold. But because of minimum length, neither higher nor lower dimensions can be deduced or detected -- neither at the Planck scale nor at larger scales.

\textbullet\       Because of minimum length and minimum time, it is  impossible to deduce or to establish the existence of any kind of space-time \textit{manifold} or of any kind of manifold with additional structure, neither at the Planck scale nor at larger scales. 
Therefore, no space-time foam, no microscopic wormholes, no diffeomorphism invariance, no fermionic coordinates, no non-commutative coordinates, no twistors, no conformal symmetry, no supersymmetry, no supergravity, no asymptotic safety, no causal fermion systems, no geometrodynamics, no T-duality, no exact UV-IR symmetry, no exact holograhic principle, no canonical quantum gravity, no continuous gauge-gravity duality, and no additional continuous or discrete space-time symmetries can be deduced, measured or confirmed.

\textbullet\       Minimum length and minimum time also imply that there are no measurable \textit{points} in space or instants in time, but that there is a Planck-scale non-locality in nature. In contrast, at larger scales, points in space and instants in time are useful \textit{approximations}. 

\textbullet\       Minimum length and minimum time imply that -- both at the Planck scale and at larger scales -- space-time structures based on \textit{discrete points} cannot be measured, detected or be proven to exist.
This includes space-time lattices, causal dynamical triangulations, spin networks, Turaev-Viro models, and graphs. 
Likewise, because of minimum length, there are no singularities in nature, of any type. The limit cube implies that they do not and cannot occur.


In short, the physics cube and the contained limit values, and in particular the minimum length and the minimum time, eliminate all descriptions of space-time and of quantum gravity that use \textit{trans-Planckian quantities}, such as descriptions based on \textit{manifolds} or on \textit{discrete points}.
The lack of any other trans-Planckian physics is also predicted.
All these predictions agree with all experiments and all observations ever performed. 
Nevertheless, the conclusions are dramatic. 
In particular, they appear to be deeply discouraging. 
Fortunately, this appearance is deceptive.

\section{From horizons to the common constituents of space and particles}

\noindent
In the 1970s, Bekenstein and Hawking showed that black holes have entropy. 
The Schwarzschild black hole is the simplest case, with an entropy given by $S/k= A/A_{\rm min}$,
where $k$ is the Boltzmann constant.
The entropy value is finite and depends on the minimum area $A_{\rm min}=4G\hbar/c^3$, the square of the minimum length. 
The Bronshtein limit cube thus implies that the factor 4 in black hole entropy is due to the factor 4 appearing in the limits of general relativity and of classical gravity.

The finite value of black hole entropy implies that gravitational horizons are made of many \textit{discrete, Planck-sized degrees of freedom.} 
This connection is explored in various approaches to quantum gravity \cite{nicolai2014quantum}; for these explorations, the limit cube turns out to be particularly helpful.

Black holes can be seen either as specific configurations of curved space or as highly concentrated matter systems.
As a consequence, \textit{both} space \textit{and} particles are made of Planck-sized degrees of freedom.
These tiny degrees of freedom of nature are the \textit{common constituents} of space and particles.
The known properties of space and particles, together with the limit cube, allow deducing the basic properties of these constituents.

\medskip

\textbullet\  Because the common constituents make up empty space, which is extended, the constituents must be \textit{extended} as well.
Because the common constituents realize the minimum length and make up particles, which are almost point-like, the constituents must have Planck radius.
However, the common constituents \textit{cannot} be of Planck size in all three dimensions. 
If that were the case, physical observables that depend on the enclosed volume would exceed the density and entropy limits set by black holes, such as Bekenstein's entropy bound \cite{bekenstein1981universal}. 
Only discrete constituents that are extended comply with the limits of the physics cube. 

\textbullet\ The common constituents \textit{cannot} be membranes extended in two dimensions
Such membranes cannot form localized structures, such as particles. 
The common constituents \textit{cannot} be bands described by an additional width parameter. 
If that were the case, black hole entropy would depend on that parameter, and not depend purely on the horizon area. 
As a consequence, the common constituents must be \textit{filiform} \cite{Carlip:2017eud,botta,ninelines}, with a Planck-scale cross section.

\textbullet\ The filiform constituents \textit{cannot} be or form loops, chain rings or connecting structures of \textit{fixed} size or shape. 
For Planck-scale sizes or shapes, observables would again exceed the density and entropy limits.
For measurably large sizes and shapes, the common constituents \textit{cannot} be fixed in shape or size, because in that case space would be anisotropic or not Lorentz-invariant -- or both.
The common filiform constituents must therefore be \textit{fluctuating}.

\textbullet\ The common constituents \textit{cannot} have ends, branches or crossings, and \textit{cannot} disappear into other dimensions. 
All these options disagree with the minimum length, with black hole entropy and with the properties of empty space. 
The common constituents must be filiform and be, apart from their Plank-scale cross section, essentially one-dimensional. 

\textbullet\ To comply with minimum length, the common constituents \textit{cannot} pass each other, but must lead to tangles, weaves and similar configurations. 
Otherwise, minimum length could not be ensured, macroscopic empty space would not be three-dimensional, and spatial curvature could not be recovered. 
Inhomogeneous configurations of the filiform constituents lead to spatial curvature. 
The filiform constituents thus act like a specific type of causal sets \cite{lucaultimo}.

\textbullet\ The common constituents \textit{cannot} have (or `carry') observable properties -- such as field intensities, mass, energy, momentum, charges, spin or other quantum numbers -- because in that case, the vacuum would have the same properties and not be empty, i.e., with vanishing observables and quantum numbers.

\textbullet\ Because the minimum length and the minimum time do limit the measurement precision of every observable and thus lead to large measurement uncertainties at Planck scales, the filiform constituents \textit{cannot} be observed individually and cannot have an equation of motion \cite{ninelines}. 
The constituents can only be described statistically.
As a consequence, no new quantum gravity effects are expected to be observable.




\textbullet\ The common filiform constituents lead to physical observables only when they form certain \textit{configurations} that are linked or tangled.
Particles, probability densities, horizons, curvature, and all physical observables are all due to specific linked or tangled configurations of the filiform constituents. 

\textbullet\ In black hole horizons, the nearly two-dimensional configurations of the filiform constituents lead to entropy and mass. 
Because of the two-dimensional configurations, for an observer at spatial infinity, mass is distributed over the horizon and not concentrated at its center \cite{ha2018external}.

\textbullet\ In flat space, specific, almost localized, fluctuating, linked or tangled  configurations of the filiform constituents lead to particles and wave functions.

\medskip

\noindent 
None of these conclusions disagrees with observations. However, filiform constituents are hard to check. Are they pure speculations? Clearly, filiform constituents are only worth considering if they imply both general relativity and particle physics.
And it appears that they do.

\medskip

\textbullet\ In the approximation of localized mass-energy distributions, the configurations of fluctuating filiform constituents of Planck radius lead to black hole entropy, black hole temperature, and black hole mass.
Using Jacobson's argument, Einstein's field equations arise \cite{jacobson, cspepan,csorigin,csindian,csbh}.
Filiform constituents exclude any extension or modification to general relativity. 
This is as observed \cite{will}.

\textbullet\ In flat space, tangled configurations of the filiform constituents lead to particles.
Specifically, \textit{rational tangles} of constituents lead to elementary and composed particles, spin, wave functions and their evolution equations.
Classifying rational tangles leads to the three particle generations and to all the elementary fermions and bosons of the standard model, including the Higgs -- and to no additional ones
\cite{cspepan,csorigin,csqed,csqcd}. 
Filiform constituents exclude elementary dark matter particles, anyons or any other additional elementary particles. This is as observed \cite{pdgnew}.

\textbullet\ In flat space, deformations of the filiform constituents lead to gauge interactions.
Classifying deformations with Reidemeister moves leads to the observed gauge groups U(1), broken SU(2), and SU(3) \cite{cspepan,csorigin,csqed,csqcd}. 
The Lagrangian of the standard model arises.
Filiform constituents exclude unified gauge groups and additional symmetries and  any modification of the standard model. This is as observed \cite{pdgnew}.

\textbullet\ In flat space, the spatial configurations of the filiform constituents with Planck radius lead to unique elementary particle mass values, unique coupling constants and unique mixing angles; their values are  close to the observed values \cite{cspepan,csorigin,csindian,csbh,csqed,csqcd}. 
So far, only rough estimates are available.
Precise numerical calculations still need to be performed.
They are predicted to agree with the data.

\medskip

In short, Bronshtein's limit cube allows deducing that nature is made of fluctuating filiform constituents of Planck radius, so-called \textit{strands}, that fluctuate in 3 dimensions. 
Rational tangles and their Reidemeister moves yield both the Lagrangian of general relativity and the Lagrangian of the standard model of elementary particle physics.
As expected from the physics cube of limits, strands predict the lack of deviations from known physics. 
No consequence deduced from tangles of strands disagrees with experiment.

\section{Conclusion and outlook}

\noindent
Using the quadruple gravitational constant $4G$ and the maximum force $c^4/4G$, Bronshtein's physics cube can be extended to exhibit an invariant limit at each of its upper corners.  
Each limit defines a theory of modern physics and is given by a Planck value, with $G$ subsituted by $4G$ everywhere.
The limit cube of physics predicts the lack of trans-Planckian physics, in agreement with all experiments so far.
The limit cube also implies that the minimum length $\sqrt{4G\hbar/c^3}$ is the single principle defining relativistic quantum gravity.
As a consequence, space-time is only effectively continuous and only effectively 3+1-dimensional.

The minimum length implies that space and particles are made of common constituents that are filiform, of Planck radius, and fluctuating.  
In separate publications, such filiform constituents, or strands, have been shown to explain the origin of general relativity and its field equations, of elementary particles, wave functions, gauge interactions and the full Lagrangian of the standard model, with unique particle masses, gauge coupling constants and mixing angles. 
Strands predict the lack of new physics, in agreement with all experiments so far.

Bronshtein's limit cube of physics thus predicts that relativistic quantum gravity, the complete description of motion, is {near} completion.
To achieve completion, 
the fundamental constants of the standard model need to be calculated to high precision.

\bigskip

\hbox to \textwidth{\hss\emph{$*\ \ \ \ \ * $}\hss} 

\newpage

\section*{Acknowledgments} 

\noindent
The author thanks Chandra Sivaram, Arun Kenath, Peter Schiller, Thomas Racey, Uwe Hohm, and Isabella Borgogelli Avveduti for fruitful discussions. %
This work was partly supported by a grant from the Klaus Tschira Foundation.

\comment{
\bigskip
\bigskip
\bigskip

\begin{quotation}
\noindent 
There is something fascinating about science. 
One gets such wholesale returns of conjecture out of such a trifling investment of fact.\\
\hspace*{1.0cm}Mark Twain, \emph{Life on the Mississippi.}
\end{quotation}
} 

\comment{
\section{Appendix -- Deriving universal gravity from maximum force and maximum speed}

Maximum force $c^4/4G$ is an energy per length.
As a consequence, a spherical area $A$ with circumference $\pi D$ can contain a maximum energy.  %
The limit is
\begin{equation}
    F_{\rm max} = \frac{c^4}{4G} = \frac{E}{A} \, {\pi D} \;\;.{\ } \label{eq:e2}
\end{equation}
Special relativity limits the acceleration $a$ of a test mass on a sphere of diameter $D$ by %
\begin{equation}
	a = \frac{c^2}{D} \;\;.  \label{eq:e3}
\end{equation}
Combining the two expressions yields
\begin{equation}
    E = \frac{c^2}{4 \pi G} \, a\, A \;\;. \label{eq:e4}
\end{equation}
This result of maximum force is a simple version of the first law of horizon mechanics \cite{Bardeen:1973gs,wald}. %

In the case of flat space, away from gravitational horizons, the relations $E=Mc^2$ and $A=4 \pi r^2$ yield
\begin{equation}
	 a= \frac{MG}{r^2} \;\;.
\end{equation}
In summary, inverse square gravity is a consequence of maximum force 
and maximum speed in the case of flat space.
} 

\bibliography{B2}

\end{document}